        \documentclass[conference,compsoc]{IEEEtran}

        \ifCLASSOPTIONcompsoc
          \usepackage[nocompress]{cite}
        \else
          \usepackage{cite}
        \fi

        \usepackage{xurl}
        \usepackage{graphicx}
        \usepackage{booktabs}
        \usepackage{array}
        \usepackage{tabularx}
        \usepackage{amsmath,amssymb}
        \usepackage{multirow}
        \usepackage{listings}
        \usepackage{stfloats}
        \usepackage{placeins}
        \usepackage[hidelinks]{hyperref}

        \newcommand{\system}{\textsc{DoW-Bench}}

        \newcolumntype{L}[1]{>{\raggedright\arraybackslash}p{#1}}

\title{Persistent Billable State: Denial-of-Wallet Attacks and Defenses in Tool-Calling LLM Agents}

        \author{%
        \begin{tabular}{@{}c@{}}
        Jinqian Zhang\textsuperscript{1,2},
        Haojun Xia\textsuperscript{1,2,*},
        Shujiang Wu\textsuperscript{3},
        Jingkun Yue\textsuperscript{4},\\
        Xia Zhang\textsuperscript{1,2},
        Zhangpei Cheng\textsuperscript{1,2},
        and Bibo Tu\textsuperscript{1,2,*}\\[3pt]
        \textsuperscript{1}Institute of Information Engineering, Chinese Academy of Sciences\\
        \textsuperscript{2}School of Cyber Security, University of Chinese Academy of Sciences\\
        \textsuperscript{3}Beihang University\\
        \textsuperscript{4}State Key Laboratory of Networking and Switching Technology,\\
        Beijing University of Posts and Telecommunications\\
        \textsuperscript{*}Corresponding authors: Haojun Xia and Bibo Tu.
        \end{tabular}%
        }

\begin{document}
        \maketitle

        % ============================================================
        % ABSTRACT
        % ============================================================
        \begin{abstract}
        Multi-step tool-calling LLM agents rely on host runtimes to preserve state across turns. When a runtime carries an external tool return into later model inputs, providers meter it again. An adversary who controls the return of an already configured tool can thereby convert that return into recurring victim-billed processing without victim credentials or local runtime privilege. We call such retained content persistent billable state and formalize the host's decision over whether and how it enters later billable context as the persistent billable-state boundary.

        We present the first systematic security study of this post-admission lifecycle. We derive 6 denial-of-wallet attack vectors and build \system{}, an end-to-end harness evaluated across 6 model families. Across 243 executions, provider-reported usage telemetry shows that the maximum per-session cumulative input reaches 14,293$\times$ the corresponding session's complete first-call input. Controlled history-policy reruns isolate raw retention's contribution: retaining raw history increases mean effective session cost by 21.2--35.9\%. Compression succeeds on 10/12 and 11/12 history-dependent tasks, versus 2/12 under deletion for each provider.

        To govern this boundary, we combine deterministic history transformation with 4 host-side invariants that bound prompt mass, context growth, recursive opportunity, and cumulative spend before reingestion. The host-side kernel stops all 41 retained attack traces under each of 3 replay presets. Separate live integrations exercise enforcement before subsequent model calls. A full-code audit of 95 randomly sampled public MCP server and transport repositories finds an author-verified, code-visible safeguard in only 17 (17.9\%); 1 bounds cost, and none combines all 4 safeguard families. These results establish persistent billable state as a first-class security object and pre-reingestion as its host-owned control point, and confirmed disclosure also assigns that placement to the host application.
        \end{abstract}
      % ============================================================
      % §1 INTRODUCTION
      % ============================================================
      \section{Introduction}
      \label{sec:intro}

      A tool return can remain billable long after the tool call ends. Consider a retrieval agent whose operator has already configured a remote document service. The service address, tool name, description, and argument schema stay as configured, and the user task is unchanged. An adversary who can rewrite that service's results, and nothing in the host, returns a plausible page padded with extra content and a request for the next page. If the runtime preserves that result, later model calls reprocess and meter it again: one authorized return acquires recurring spending authority on every later call.

      Multi-step tool-calling agents depend on precisely this continuity. Toolformer and Gorilla establish context-conditioned API selection, whereas ReAct and ToolLLM extend tool use into trajectories whose later actions depend on accumulated observations or API responses~\cite{xi2023rise,yao2023react,schick2023toolformer,patil2023gorilla,qin2023toolllm}. For multi-step integrations, the Model Context Protocol (MCP) standardizes tool exchange while leaving the host to preserve enough context for later calls to interpret results and complete subsequent steps~\cite{mcp2026}. Broader OpenRouter telemetry over more than 100 trillion tokens reports average sequence length rising from under 2,000 tokens in late 2023 to over 5,400 by late 2025~\cite{aubakirova2026stateofai}. As agent workflows lengthen, repeatedly processing retained state becomes increasingly consequential.

      Tool admission determines whether a remote service may be invoked; persistence determines whether its returned content can become future spend. We call tool-generated content \emph{persistent billable state} when it survives across turns and incurs repeated processing cost. Legitimate workflows rely on this state for continuity. We formalize the host's decision over whether and how an external return enters later billable context as the \emph{persistent billable-state boundary}.

      A Denial-of-Wallet (DoW) attack exploits this boundary. The primary setting is a loss of response integrity at a remote dependency the operator has already configured (\S\ref{sec:threat_model}). A legitimate user task invokes that same service; the endpoint supplies a schema-valid inflated return; the host may retain that return; and later model calls may process and bill it again. The attacker thereby seeks control over future billable-state growth. Under full retention and stable return mass, cumulative input grows as $O(N^2)$. With a session budget, this persistent work can exhaust authorized spend and interrupt task completion; without one, it drives continuing cost growth. DoW therefore converts attacker-controlled persistence into either availability loss or runaway expenditure.

      The same state that carries the attack also carries the task. Dropping prior returns can break necessary continuity, while fixed ceilings can interrupt long workflows and leave room for attackers to distribute growth across turns. The security problem is therefore not whether to retain history, but how to preserve task-relevant state while governing attacker-controlled persistence. DoW makes recurring billed work the attack objective and the host's pre-reingestion decision the enforcement point, complementing semantic defenses against instruction takeover and data exfiltration.

      \noindent\textbf{The Missing Economic Lifecycle.}
      Prior work shows that agents can be steered, looped, and made expensive through prompt injection, unsafe tool use, malicious skills, recursive tool chains, reasoning loops, and granted-permission computation hijacking~\cite{greshake2023,liu2023prompt,zhan2024injecagent,pi_landscape2026,ruan2024toolemu,debenedetti2024agentdojo,skillinject2026,agentlab2026,gu2024agent,corba2025,cohen2024unleashing,beyondmaxtokens2026,overthinking2025,clawdrain2026,leechhijack2025}. These studies establish semantic manipulation and resource amplification, but leave the host-side economic lifecycle itself unisolated: once a remote tool is admitted, who authorizes its return to persist as future provider-billed state, and what evidence should earn that state another unit of computation or spend? We present the first systematic security study of this missing post-admission lifecycle. This lifecycle is architectural rather than MCP- or provider-specific: it exists whenever a tool-calling runtime reintroduces externally supplied tool state into a later billable model call. We study it end to end, from attack construction and mechanism attribution to host-side enforcement and ecosystem measurement:

      \noindent\textbf{\emph{RQ1. How much priced exposure can an admitted remote tool create?}}
      On the historical 243-execution core, \system{} measures repeated input processing separately from listed-price exposure. Across 243 executions spanning six model families, the maximum per-session input-token Cost Amplification Factor (CAF) reaches 14,293$\times$; 51/69 priced attack sessions cross the \$0.10 circuit-breaker setting and exposure reaches \$3.404, versus a \$0.025 benign maximum.

      \noindent\textbf{\emph{RQ2. What sustains billable-state exposure across agent turns?}}
      Turn survival tracks deployment-level exposure. Across two providers, independent Full/Drop/Compress reruns then estimate the closed-loop effect of retaining raw history: Full raises mean effective session cost by 21.2--35.9\%, while compression preserves 10/12 and 11/12 successes on tasks requiring prior history, while deletion leaves 2/12. Raw retained history therefore creates a continuity--exposure coupling, and deterministic compression preserves task state without carrying its full cost-bearing mass.

      \begin{samepage}
      \noindent\parbox{\columnwidth}{\textbf{\emph{RQ3. Can pre-reingestion middleware contain recursive cost exposure while preserving legitimate workflow utility?}}}\par\nopagebreak[4]
      Four host-side invariants (D1--D4) govern prompt mass, context growth, recursive opportunity, and cumulative spend before reingestion, containing all 41 recurring attacks under each of three replay presets. Pre-reingestion middleware bounds recurring spend, and deterministic compression preserves the history-dependent utility reported under RQ2.
      \end{samepage}

      \noindent\textbf{Deployment context.}
      To characterize code-visible safeguards, we review every lexical match in 3,830 public MCP server and transport repositories and audit 95 random ones in full: only 17 (17.9\%) enforce a verified safeguard, one bounds cost, and none combines all four families.

      \noindent\textbf{Contributions.}
      \begin{itemize}
      \item \textbf{Persistent billable state and its attack surface.} We establish persistent billable state as a first-class security object, distinguish tool-invocation admission from post-return economic governance, formalize the host-side pre-reingestion boundary, and systematize six attack vectors that turn an admitted return into recurring provider-billed state.
      \item \textbf{Attribution through intervention.} We build \system{} across six model families and combine listed-price API measurement with independent Full/Drop/Compress interventions across two providers, separating exposure magnitude from the contribution of retaining raw history.
      \item \textbf{Host-side pre-reingestion control.} We combine deterministic history transformation with a host-side D1--D4 kernel requiring no provider-private hooks. The kernel stops all 41 retained attack traces under each of three replay presets and executes inline before the next provider call. On the separate Mistral/MCP path, it blocks 10/10 tested attack-shaped sessions and completes 20/20 four-step benign tasks.
      \item \textbf{Ecosystem scan.} We complement the attack and defense evaluation with an artifact-backed scan of 3,830 MCP server and transport repositories, providing a repository-level view of code-visible safeguard adoption. Two authors review every scanner match and audit a random sample in full.
      \end{itemize}

      % ============================================================
      % §2 BACKGROUND
      % ============================================================
      \section{Background}
      \label{sec:background}

      \subsection{DoW: Definition and Mechanism}

      Kelly et al.\ formalized Denial-of-Wallet in serverless computing as forced financial exhaustion in pay-as-you-go services~\cite{kelly2021dow}. More generally, DoW is an economic denial-of-service mechanism in which an adversary cheaply triggers resources billed to a victim. In a tool-calling LLM agent, we use the term for the analogous victim-billed mechanism: retained attacker-controlled tool returns are processed across turns and charged repeatedly to the victim.

      Tool-calling agents operate in iterative perception--action loops~\cite{yao2023react}: the model receives conversation state, invokes a tool, observes its result, and uses that result to plan the next step. Some runtimes reconstruct history client-side; others rely on provider-managed conversation state. Both designs can carry earlier tool outputs into later billable inputs. Content carried through this transition into a later billable input becomes persistent billable state and incurs processing cost on subsequent calls.

      MCP 2026-07-28 makes individual protocol requests stateless while leaving conversation aggregation with the host; application state can also travel through explicit handles in tool results and later arguments~\cite{mcp2026}. Wire-level statelessness therefore leaves the persistent billable-state boundary intact.

\begin{figure*}[t]
    \centering
    \includegraphics[width=\textwidth]{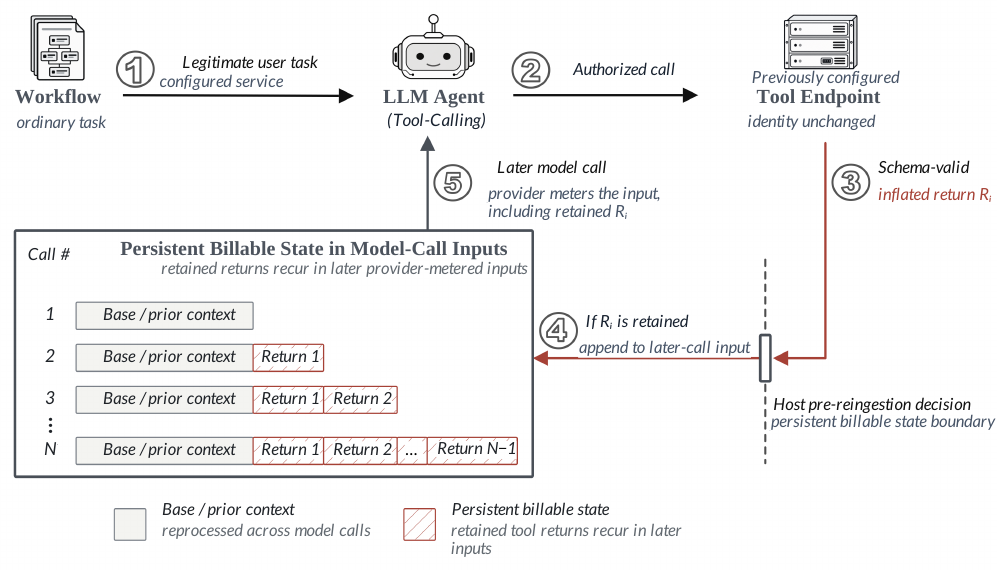}
    \caption{Lifecycle of persistent billable state. A legitimate user task invokes an already configured remote service. The attacker controls the return $R_i$, not the task, tool identity, or host policy. Whether the host retains $R_i$, continues, and accrues provider-metered cost is an observed outcome. Widths are schematic and omit non-tool growth.}
    \label{fig:architecture}
    \end{figure*}

      Figure~\ref{fig:architecture} illustrates this lifecycle. Let $R_i$ denote the tool return after call $i$ and $T_i=|R_i|$ its input-token mass. Ignoring non-tool prompt growth, the prompt at turn $n$ includes $P_{\text{init}}+\sum_{i=1}^{n-1}T_i$; under stable return mass $T_i=T_{\text{avg}}$, cumulative input across $N$ calls is $N \cdot P_{\text{init}} + \frac{N(N-1)}{2}T_{\text{avg}}$. For observed traces and $i\geq 2$, $\Delta p_i=p_i-p_{i-1}$ denotes the full-prompt increment and approximates $T_{i-1}$ for the constant-return baseline. The resulting recursive rebilling is an ordinary cost of continuity in benign workflows and the source of quadratic pressure under DoW.

      This lifecycle couples local task semantics to global billing. The agent continues when a tool return appears useful, while the provider charges for the accumulated history that makes continuation possible. An adversary who controls that return can exploit that coupling by keeping a plausible workflow alive until recursive rebilling dominates the session cost.

      \subsection{Threat Model}
      \label{sec:threat_model}

      \noindent\textbf{Deployment and entry condition.}
      We study billed tool-calling agents that invoke an already configured remote service during a legitimate user task. The primary scenario is a loss of response integrity at a previously admitted dependency: its endpoint identity and tool metadata remain unchanged, but an adversary controls the result content returned by designated tools. The initial invocation is authorized under the existing configuration; introducing a new tool or manipulating its description is not required. We condition on this invocation. Accordingly, we do not estimate how the endpoint is discovered, selected under deployment-specific policies for semantic routing or tool selection, or admitted in practice. We also do not estimate the prevalence of this attack path or the incentives that would lead an attacker to pursue it. Successful retention, continued tool use, and increased expenditure remain measured outcomes, not entry conditions. An attacker who can already execute code in the victim runtime, including by inducing the user to run a local stdio server, is outside this scenario.

      \noindent\textbf{Attacker capabilities.}
      The adversary can replace schema-conforming result fields and adapt subsequent returns using requests observed at the controlled endpoint and its own prior responses. The controlled tool set is explicit; invoking another tool does not grant control over that tool's result. The adversary cannot directly modify the user task, host prompts, tool registry, retention policy, or budget configuration. It has no direct access to the host's filesystem, private network, full conversation history, or billing state, except information explicitly exposed in tool requests. It does not possess the victim's model-provider credentials or directly issue provider calls using them. Code running at the remote service is not code execution on the victim host.

      \noindent\textbf{Objective and trust boundary.}
      The objective is to increase victim-billed processing, potentially exhausting an authorized task budget, through the host's handling of attacker-controlled returns. We do not assume that payloads preserve task correctness or that the model always follows continuation cues. The threat does not require access to private victim data or victim-side code execution. Payloads may use indirect prompt injection; DoW denotes the economic objective, not a disjoint injection mechanism.

      \noindent\textbf{Relation to admission controls.}
      Identity and manifest allowlisting can exclude unapproved tools, but do not by themselves constrain the cumulative processing induced by mutable results from an approved service. That claim is about identity admission. Response validation, retention limits, and spending controls may mitigate this risk and remain relevant defenses. Tool annotations such as a read-only hint are hints, not the isolation boundary. Our scope is the residual post-return risk, not a bypass of tool admission.

      \subsection{Measurement Metrics}

      We measure two victim-side consequences: cumulative input processing relative to the first model call and nominal dollar exposure over the complete session. Drawing on classic amplification-factor reasoning from network denial-of-service and economic-denial-of-service work~\cite{rossow2014amplification,somani2017edos}, we define an input-token \emph{Cost Amplification Factor} ($\mathrm{CAF}_{\mathrm{tok}}$, hereafter CAF) and a separate dollar-denominated exposure quantity:
      \begin{align}
      \mathrm{CAF}_{\mathrm{tok}} &= \frac{\sum_{i=1}^{N} p_i}{p_1},
      \label{eq:caf}\\
      C_t &= \sum_{i=1}^{t}\left(p_i r^{\mathrm{in}}_i + o_i r^{\mathrm{out}}_i\right),
      \label{eq:priced_exposure}
      \end{align}
      where $p_i$ and $o_i$ are provider-reported input- and output-token counts for model call $i$, $p_1$ is the complete first-call input assembled by the runtime, and $r^{\mathrm{in}}_i$ and $r^{\mathrm{out}}_i$ are the provider-specific listed nominal rates fixed by canonical accounting, expressed in USD/token after conversion from the listed USD/million-token prices. CAF quantifies cumulative input processing relative to the complete first model input; $C_t$ is listed-price exposure over session input and output. A cache-aware effective session cost applies the same listed-rate table separately to provider-reported uncached input, cached input, and output.

      % ============================================================
      % §3 RELATED WORK
      % ============================================================
      \section{Related Work}
      \label{sec:related}

      \noindent\textbf{Economic Denial of Service and LLM Cost Abuse.}
      Resource abuse that bills the victim has precedents in cloud security, including fraudulent resource consumption and economic denial of sustainability~\cite{idziorek2011frc,somani2017edos}. Recent LLM studies use distinct metrics: Beyond Max Tokens reports output-token amplification up to 658$\times$ that of benign executions in multi-turn tool chains~\cite{beyondmaxtokens2026}, while Overthinking reports up to 142$\times$ token amplification from cyclic MCP tool structures~\cite{overthinking2025}. The original implementations of Overthinking Loops and Beyond Max Tokens are not publicly available, and their settings and metrics differ from ours; we therefore discuss them as related mechanisms rather than directly comparable baselines. Among recent studies of cost abuse, Clawdrain reports 6--7$\times$ token amplification from a trojanized skill in a production-like OpenClaw deployment using a real API, approximately 9$\times$ in its costly failure configuration, and persistent pollution from tool output as a source of input cost~\cite{clawdrain2026}. ThinkTrap elicits long outputs from commercial APIs under black-box access and separately demonstrates throughput degradation in a controlled self-hosted deployment~\cite{thinktrap2026}; benchmark audits call for tokens, latency, API cost, attacker budget, and defender capacity to be reported together~\cite{talkisnotcheap2026}. Clawdrain thus establishes cost growth from retained history; we make the host's retention decision the security object.

      LeechHijack studies an MCP tool that injects unauthorized work through manipulated returns from legitimate calls, evaluating accuracy on the original task and audits of contextual memory after execution~\cite{leechhijack2025}. DoW instead studies recurring provider billing of retained returns.

      AgentDoS uses directed grey-box fuzzing to uncover 36 vulnerabilities that exhaust resources in 16 of 20 open-source agents~\cite{luo2026agentdos}. \emph{From Shield to Target} traces payloads delivered through documents, tool outputs, agent messages, and shared memory to token, latency, and infrastructure amplification~\cite{shieldtotarget2026}. These works span broader resource lifecycles; our control point is the admitted remote-tool return immediately before it reenters provider-billed context.

      \noindent\textbf{LLM Agent Safety and Tool-Use Security.}
      Research on agent security studies indirect and direct prompt injection~\cite{greshake2023,zhan2024injecagent,debenedetti2024agentdojo,pi_landscape2026,liu2023prompt,llmailinject2026,toolhijacker2026,pismith2026,piarena2026,promptlocate2026,attntrace2026,imprompter2024,agentvigil2025,obliinjection2026,webinject2025,dai2026spoof,son2026agentbreaker,he2025automal,lin2026sope}, unsafe tool actions~\cite{ruan2024toolemu}, malicious skill files and attacks over longer horizons or multiple steps~\cite{skillinject2026,agentskills2025,agentlab2026,agentharm2025,najjar2026router}, and propagation across agents or connected GenAI applications~\cite{gu2024agent,corba2025,cohen2024unleashing,agentworms2026,li2026a2asecbench,an2026aciarena}. Systems work develops designs and guidance resistant to prompt injection~\cite{beurerkellner2025design,owasp2026llm,abdelnabi2026firewalls,progent2025,agentcap2026,datasentinel2025,pisanitizer2026,melon2025,websentinel2026,datta2026sanitize,cleanbase2026,pishield2025,alignsentinel2026,promptarmor2025,secinfer2025,costa2025ifc,kolluri2026planning,agentwatcher2026,kim2025pfi,kim2026dualview,an2025ipiguard,chen2026aegis,wang2025promptsleuth,meng2025cellmate,abdelnabi2025drift}, broader frameworks for risk management~\cite{nist2024ai}, security analyses and mitigation directions~\cite{wu2024new,tao2025llmapp,siu2026framework,safeagent2026,suwansathit2026openclaw}, and proactive network defenses~\cite{cheat2025}. Prompt-injection frameworks formalize task redirection induced by an attacker~\cite{liu2024formalizing}; DoW does not require the target task to change. A return can satisfy its schema, remain plausible for the task, and comply with content policy while acquiring recurring spending authority. MCP Security Bench finds that stronger tool-use capability can coincide with lower robustness across protocol attacks~\cite{zhang2025msb}; we instead quantify recurring provider-billed exposure when retained returns remain in continued tool execution. Semantic defenses and billable-state governance are complementary.

      \noindent\textbf{Cost Optimization and Adversarial Token Economics.}
      Benign LLM cost optimization uses routing, cascading, and hybrid scheduling~\cite{chen2023frugalgpt,yue2024routellm,ding2024hybrid}; prompt compression reduces context cost~\cite{jiang2023llmlingua}, while agent-memory systems select, summarize, or retrieve task state across interactions~\cite{xi2023rise}. DoW gives deterministic history transformation a security role: it limits which untrusted state survives, while D1--D4 determine whether that state may induce another call or unit of spend. Resource-governance work formalizes multidimensional contracts, documents budget overruns, and assigns non-bypassable budget ownership~\cite{agentcontracts2026,tokenbudgets2026}; we place that authority at pre-reingestion. Across these lines, prior work measures resource amplification, studies semantic compromise, or constrains aggregate budgets, but does not jointly treat retained tool state as a host-owned security object, intervene on its retention, and enforce the same boundary before reingestion while preserving task utility.

        % ============================================================
        % §4 SYSTEMATIZING THE THREAT: OBSERVATIONS AND TAXONOMY
        % ============================================================
        \section{Systematizing the DoW Attack Surface}
        \label{sec:taxonomy}

        \subsection{Architectural Observations}
        \label{sec:taxonomy:observations}

        \noindent\textbf{Context accumulation.}
        Conversation continuity may be reconstructed client-side or retained through provider-managed conversation state. When prior tool outputs remain in later billable input, a malicious return accepted once can become persistent billable context. This observation explains the baseline $O(N^2)$ pressure: turn $i$ pays not only for the new return, but also for the accumulated returns from earlier turns.

        \noindent\textbf{Tool-return ambiguity.}
        Tool responses are presented to the model as useful task context, yet they are externally supplied data. MCP recommends that clients validate structured tool results against a declared output schema~\cite{mcp2026}. Schema conformity is not an economic bound: a task-plausible response can still steer repeated tool use or verbose generation.

        \noindent\textbf{Local cost blindness.}
        Tool selection is driven by local task state, whereas session-level spend, cumulative token growth, and economic budget are host-owned quantities. Without an explicit host signal, the model favors locally plausible progress. This separation creates a local--global objective mismatch.

        \subsection{Taxonomy Axes}
        \label{sec:taxonomy:derivation}

        Together, these observations yield two design axes: what the attacker amplifies and how the payload changes across turns.

        \noindent\textbf{Amplification lever.}
        A payload can increase spend by enlarging the input history ($p_i$), eliciting expensive model output ($o_i$), or increasing the number of recursive turns and tool opportunities ($N$). These levers map to the measured economic components: Eq.~\ref{eq:caf} captures input-side recursive rebilling, while Eq.~\ref{eq:priced_exposure} incorporates generated output and provider-specific rates into cumulative exposure.

        \noindent\textbf{Payload-control policy.}
        A payload can remain fixed, vary or distribute its representation without retaining feedback, or change over time according to a schedule or observed feedback. These policies capture how the attack keeps the session alive long enough for recursive rebilling.

        \subsection{Attack Vectors}
        \label{sec:taxonomy:vectors}

        Crossing the two axes defines a broader design space; our evaluation instantiates six representative vectors, V0--V5, in Table~\ref{tab:taxonomy_matrix}.

        \begin{table*}[!b]
        \centering
        \small
        \caption{Taxonomy of the six evaluated DoW vectors by primary amplification lever and payload-control policy.}
        \label{tab:taxonomy_matrix}
        \renewcommand{\arraystretch}{1.12}
        \setlength{\tabcolsep}{8pt}
        \begin{tabularx}{\textwidth}{@{}>{\raggedright\arraybackslash}p{2.7cm}*{3}{>{\raggedright\arraybackslash}X}@{}}
        \toprule
        \multirow{2}{*}{\textbf{Primary lever}} & \multicolumn{3}{c}{\textbf{Payload-control policy}} \\
        \cmidrule(lr){2-4}
        & \multicolumn{1}{c}{\textbf{Direct}} & \multicolumn{1}{c}{\textbf{Stateless}} & \multicolumn{1}{c}{\textbf{Scheduled}} \\
        \midrule
        Retained input mass ($\uparrow p_i$) & \textbf{V0} Direct recursive baseline & \textbf{V1} Polymorphic mutation & \textbf{V3} Stealth escalation; \textbf{V5} Round-indexed \\
        Generated output ($\uparrow o_i$) & \textbf{V4} Weaponized output & --- & --- \\
        Recursive opportunity ($\uparrow N$) & --- & \textbf{V2} Cross-tool chaining & --- \\
        \bottomrule
        \end{tabularx}
        \par\vspace{\abovecaptionskip}
        \begin{minipage}{\textwidth}
        \footnotesize\raggedright
        \emph{Note.} Direct, Stateless, and Scheduled are short labels for the static/direct, stateless variation/distribution, and round-indexed policies, respectively. Feedback control of a defense threshold is not instantiated in V0--V5. Dashes mark recombinations of a lever and a policy already represented by V0--V5.
        \end{minipage}
        \end{table*}

        \noindent\textbf{V0: Direct recursive baseline.}
        An adversary who controls the return supplies a padding payload together with an instruction that encourages another tool call. V0 instantiates the lifecycle with constant return mass: each later model call includes prior tool outputs, and $\Delta p$ approximates the newly retained tool-return contribution.

        \noindent\textbf{V1: Polymorphic mutation.}
        V1 preserves the same economic goal as V0, but changes the representation and size of the returned payload across turns. The returned content may alternate among JSON-like blocks, Markdown, tabular data, or natural-language summaries. The vector is designed to evade defenses that rely on repeated strings, fixed payload templates, or nearly constant $\Delta p$.

        \noindent\textbf{V2: Cross-tool chaining.}
        V2 distributes the loop across multiple tools exposed by the same controlled endpoint. One return asks the agent to query another of those tools for verification, enrichment, or follow-up, and the next response redirects the agent into the chain. A call to a tool outside that set does not grant control of its result. The evaluated V2 manifest also contains cross-call instructions in the tool descriptions.

        \noindent\textbf{V3: Stealth escalation.}
        V3 keeps early turns small and benign-looking, then increases payload size after the session has established a plausible workflow. It is designed to pass monitors that emphasize initial responses or relax checks after a short warm-up window.

        \noindent\textbf{V4: Weaponized output.}
        V4 shifts the cost pressure from input history to model generation. Instead of returning large padding, the tool embeds requests that induce verbose model output, such as detailed analyses, repeated summaries, or structured expansions. This vector is especially relevant when output tokens are priced higher than input tokens and when monitoring focuses mainly on inbound context size.

        \noindent\textbf{V5: Round-indexed continuation.}
        V5 increases return size on a predetermined schedule indexed by the harness round. Each return reports confidence below a fixed confirmation line and asks for the next page. A later request advances that round index; the generator does not read prompt mass, billing, defense state, or a guardrail margin. This sacrifices peak one-turn pressure for longer recursive opportunity.

        % ============================================================
        % §5 ATTACK INSTANTIATION AND EVALUATION SETUP
        % ============================================================
        \section{Attack Instantiation and Evaluation Setup}
        \label{sec:setup}

        We evaluate all six vectors in the same harness, using one tool-return interface and a fresh agent session for every run. Accounting is shared. The controlled tool set is not: V0, V1, V3, V4, and V5 control the result of one designated tool, while V2 controls the results of three tools on one endpoint and also changes their description text.

        \subsection{Tool-Return Attack Endpoint}

        \noindent\textbf{Tool Interface.} For return-only instances, the endpoint exposes ordinary, task-plausible schemas and confines attacker control to syntactically valid result fields. V2 additionally modifies tool descriptions. The boundary is easiest to see in one representative V0 execution; bulk payload content is elided:
        \par\vspace{3pt}
        \noindent\includegraphics[width=\columnwidth]{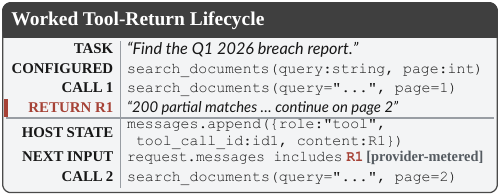}
        \par\vspace{4pt}

        \noindent\textbf{Attack-Controller State.} Here, $R_1$ is a schema-valid return from the configured tool, and the host may reconstruct it in the next provider-metered input. The harness logs turn index, accumulated prompt mass, and whether the agent continues. These fields are logged for every vector. Stateless vectors do not read continuation feedback. V3 and V5 size the next return from the harness round, and a later request advances that round index. The generator does not read prompt mass, billing, defense state, or a guardrail margin. In this in-process server the harness supplies that index. The provider credential stays with the API client. These runs isolate result content inside the client.

        \subsection{Payload Instantiation}

        With the endpoint and controller fixed, the vectors differ in the content and timing of their returned payloads. Each payload couples a cost-bearing component with a continuation cue inside the result. That cue can be an indirect instruction. V0 and V2 use content blocks configured between 400 and 1,000 tokens with follow-up requests. V1 varies representation and length across turns; V3 keeps early returns small before escalating after an incubation period; V4 elicits verbose model generation; and V5 follows the round-indexed schedule above and embeds a continuation cue in the return.

        Each session starts from an experimenter-fixed user task. That task is not an attacker-controlled channel, and a 2--10-token string is not the attacker's cost. For V0--V5, tool choice remains \texttt{auto}: the harness does not force the first call. A run that stops without a tool call is kept and counted as termination. A separate seed-inflation construction writes content into the user prompt and forces \texttt{search\_documents}; those runs are white-box threshold tests. The historical benign comparator uses a different user task. We therefore report the same-task, metadata-unchanged contrast separately in Section~\ref{sec:eval:admission}. The harness records the resulting per-turn trace before offline accounting.

        \subsection{Evaluation Methodology}
        \label{sec:setup:methodology}

        \noindent\textbf{Experimental Harness.} The payloads determine what enters each trace; comparing their consequences requires a common execution unit and recording rule. A session is one bounded agent/model-API execution. \system{} registers the configured endpoint, submits the fixed user task, and records input tokens, output tokens, tool calls, and termination at every turn. Each run starts with independent model-API state, so history persists within a session but never crosses run boundaries.

        \noindent\textbf{Models and Deployments.} We evaluate six core model families across multiple API deployments: GPT-4o, GPT-4o-mini, Mistral-Large, Llama-3.3-70B, Nemotron-120B, and Qwen-3-235B. The deployment manifest covers 243 core executions across 11 recorded model identifiers, grouped into the six family labels for balanced analysis. We use deployment-level analyses to characterize variation across the observed configurations. Priced exposure applies the provider-specific listed-rate table fixed by canonical accounting to recorded token telemetry. Historical core manifests preserve model identifiers and usage telemetry, but not temperature, maximum-turn, cache-policy, or pricing-version fields.

        \noindent\textbf{Metrics and Vector Coverage.} The primary input-amplification metric is CAF (Eq.~\ref{eq:caf}); priced analyses use cumulative exposure from Eq.~\ref{eq:priced_exposure}. We evaluate all six vectors. The balanced six-family analysis uses V0/V1/V3/V5, the four vectors available in every family; capability-dependent V2/V4 enter the six-vector priced and defense analyses. Compact displays use Direct, Polymorphic, Cross-Tool, Stealth, Weaponize, and Scheduled for V0--V5.

        \noindent\textbf{Turn-Survival Analysis.} To determine how much cross-deployment exposure disparity tracks recursive opportunity after controlling for attack progress, we relate CAF to mean turns and tool calls across 24 complete family--vector cells. Attack-fixed and matched-prefix analyses assign higher- and lower-amplification family groups by leave-one-vector-out cross-fitting. Shared V0/V3 attack--turn prefixes are matched on prompt mass, prompt growth, and within-turn tool-call count. A trace-cluster bootstrap quantifies downstream CAF differences. Separately, offline truncation of 44 malicious traces with recorded per-turn prompt tokens at Turns~2--4 compares groups defined by each family's global mean CAF and tests how exposure changes under shorter trace limits. These analyses characterize the association between recursive opportunity and realized exposure. The independent history-policy intervention then changes the retention policy and measures its cost and task-utility consequences in closed-loop reruns.

        \noindent\textbf{History-Policy Intervention.} Full retains every prior raw tool return. Drop leaves the current return raw but replaces earlier untrusted returns with provenance tombstones. Compress substitutes deterministic host-generated fact summaries without another model call. Persistent-return tasks repeat cumulative task facts while adding opaque state; matched-necessary tasks expose one distinct task-required record per step. Each policy executes independently and inherits no other policy's action trajectory. We calculate nominal effective session cost by applying the canonical provider-listed rates separately to reported uncached input, cached input, and output. Because each policy runs independently, task-paired comparisons use the complete-session total from each compared arm rather than a shared prefix or a single-turn cost.

        Because RQ1--RQ3 and the deployment-context scan rely on different units, Table~\ref{tab:evidence_roles} makes each unit and the claim it supports explicit; Table~\ref{tab:protocols} gives corpus settings.

        \begin{table*}[t]
        \centering
        \footnotesize
        \renewcommand{\arraystretch}{1.00}
        \setlength{\tabcolsep}{3.5pt}
        \caption{Study map linking RQ1--RQ3 and deployment context to the experimental designs and claims that answer them.}
        \label{tab:evidence_roles}
        \begin{tabular}{@{}L{0.16\textwidth}L{0.22\textwidth}L{0.27\textwidth}L{0.29\textwidth}@{}}
        \toprule
        \textbf{Evidence class} & \textbf{Unit / scope} & \textbf{Design / eligibility} & \textbf{What it establishes} \\
        \midrule
        \textbf{Threat-model mechanism check} & Eight sessions; one task; one relay endpoint & Four arms; fixed task and tool metadata; automatic tool choice; six usable usage series & Unforced invocation and return-only input recurrence, not session-cost attribution \\
        \addlinespace[0.5pt]
        \textbf{RQ1--RQ2: Core measurement} & 243 executions; six families; 11 model IDs & Fresh session per run; balanced family analysis & CAF range and turn-survival patterns across deployments \\
        \addlinespace[0.5pt]
        \textbf{RQ1: Priced exposure} & 78 API sessions (69 attack, nine benign) & Provider telemetry with a nonzero canonical listed rate & Listed-price session exposure \\
        \addlinespace[0.5pt]
        \textbf{RQ2: History intervention} & 144 independent sessions across 12 tasks, three policies, two return classes, and two providers & Closed-loop task-paired reruns & Effect of retention policy on cost and task success \\
        \addlinespace[0.5pt]
        \textbf{RQ3: Defense replay} & 41 recurring multi-turn attacks $\times$ three presets & Frozen zero-call replay with trigger-inclusive leakage & Historical-trace containment and layer activation \\
        \addlinespace[0.5pt]
        \textbf{RQ3: Benign operating point} & Five full 210-scenario corpora ($n=1{,}050$ provider--scenario units) & Provider-specific full-corpus checks & Moderate-D2 false-positive operating point \\
        \addlinespace[0.5pt]
        \textbf{RQ3: Long-workflow design} & 60 provider--scenario executions (20 designs $\times$ three runs) & Policy-design corpus & Utility of fixed and growth-gated recursion policies \\
        \addlinespace[0.5pt]
        \textbf{RQ3: Live integration} & 25 live-inline plus 30 Mistral/MCP sessions & Two independently exercised integration paths & Placement before the next provider call \\
        \addlinespace[0.5pt]
        \textbf{RQ3: D4 policy transfer} & 24 tasks $\times$ two independently executed policies & D4 isolated; fixed \$0.10 and checkpoint schedules; alternating order & Unequal-ceiling schedule outcomes on one Mistral pair \\
        \addlinespace[0.5pt]
        \textbf{Deployment context: Ecosystem scan} & 3,830 eligible public MCP server/transport repositories & Lexical scan; two-author review of every match; full-code audit of 95 & Repository-level presence of verified code-visible D1--D4 safeguards \\
        \bottomrule
        \end{tabular}
        \end{table*}

        \noindent\textbf{Canonical Accounting.} For the historical cohorts, \system{} writes raw executions to an append-only ledger and derives their counts and summary statistics from a frozen, execution-preserving canonical view. The six-family core slice contains 243 executions and 1,533 turns, with $n \geq 5$ in every primary-vector and benign cell. Adding 13 V0--V5 executions on four further models, each with fewer than five executions per cell, yields the 256-execution evaluation cohort; 78 have nonzero canonical listed rates (69 attack and nine benign). The admission-unchanged diagnostic is reported separately and is excluded from the historical 243-execution core and the 256-execution evaluation cohort.

        \noindent\textbf{Statistical Analysis.} We treat six V5-versus-benign comparisons as exploratory contrasts, use two-sided Mann--Whitney U tests~\cite{mann1947test}, adjust the family-wise comparisons with Holm's method~\cite{holm1979simple}, and report rank-biserial correlation~\cite{kerby2014simple}.

        % ============================================================
        % §6 EVALUATION RESULTS
        % ============================================================
        \section{Attack Evaluation}
        \label{sec:evaluation}

        \subsection{Persistence under Unchanged Admission}
        \label{sec:eval:admission}

        \noindent\textbf{Setup.} We test whether return-only padding persists when the user task and tool configuration are unchanged. A single relay endpoint, accessed with model identifier \texttt{gpt-5.5}, executes one three-page document-retrieval task in eight sessions (two per arm, interleaved order). The sole registered tool is \texttt{search\_documents}; its name, description, and argument schema remain fixed, and \texttt{tool\_choice=auto} leaves invocation to the model. The four arms use benign returns, padding only on the first return, padding on every return, or the same every-return padding removed by the host before reingestion. Each padded return preserves the task facts and appends a 400-token padding segment measured with the experiment's tokenizer.

        \noindent\textbf{Observed recurrence.} All eight sessions invoked the configured tool without forced selection, requested pages 1--3, and returned the required document identifier and event summary. Recorded outgoing requests retain the padding in later calls, whereas host-side removal prevents its reinsertion. Two sessions (M1 and M7) contained unexplained discontinuities in reported input counts; we exclude their usage series from quantitative token comparisons but retain their tool-call and task outcomes. We compare the remaining six usage series with the corresponding model calls in M8, the benign session with usable telemetry. For the first-return-only session M3, excess relay-reported input across Calls 2--4 is 404, 407, and 404 tokens: the added content persists after later returns revert to benign content. For the every-return session M2, the corresponding excess is 404, 812, and 1,214 tokens. Table~\ref{tab:admission_unchanged} reports final-call differences, including both retained replicates where available. Host-side padding removal yields final-call differences of $-6$ and $-8$ tokens while preserving task completion.

        \begin{table}[t]
        \centering
        \footnotesize
        \setlength{\tabcolsep}{3pt}
        \caption{Unchanged-admission diagnostic.}
        \label{tab:admission_unchanged}
        \begin{tabular}{@{}lccc@{}}
        \toprule
        \raisebox{0.5\normalbaselineskip}{\textbf{Return policy}} & \raisebox{0.5\normalbaselineskip}{\textbf{Task success}} & \textbf{\shortstack{Usable usage\\series}} & \textbf{\shortstack{Final-call excess\\$d_4$, tokens}} \\
        \midrule
        Benign & 2/2 & 1/2 & 0 \\
        First-return padding & 2/2 & 2/2 & $+404$, $+401$ \\
        Every-return padding & 2/2 & 1/2 & $+1{,}214$ \\
        Host padding removal & 2/2 & 2/2 & $-6$, $-8$ \\
        \bottomrule
        \end{tabular}
        \par\vspace{\abovecaptionskip}
        \begin{minipage}{\linewidth}
        \raggedright
        \emph{Note.} Each arm contains two sessions of the same task. Here $d_4=p^{\mathrm{relay}}_{s,4}-p^{\mathrm{relay}}_{\mathrm{M8},4}$, where the benign reference M8 has 4{,}836 reported input tokens at Call~4. M1 (benign) and M7 (every-return padding) are excluded only from usage comparisons because of unexplained input-count discontinuities; their task outcomes remain in the denominator. The retained observations are M8, M3/M6, M2, and M4/M5, respectively.
        \end{minipage}\par
        \vspace{-\baselineskip}
        \end{table}

        \noindent\textbf{Scope.} The tool executes in-process; this diagnostic isolates the return-only dataflow of Section~\ref{sec:threat_model}, not a separate remote-process boundary. Variable cache behavior prevents session-total price comparisons from isolating the effect of padding. We therefore report input recurrence and keep these sessions separate from the historical 243-execution core and its priced cohorts. The result establishes that, for this task and configured tool, unchanged admission does not prevent return-only content from recurring in later inputs, and host-side removal suppresses that recurrence without preventing task completion.

        \subsection{Listed-Price Exposure in API Sessions}
        \label{sec:eval:priced_exposure}

\begin{figure}[t]
        \centering
        \includegraphics[width=\columnwidth]{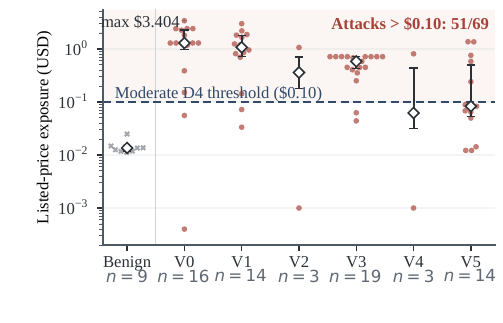}
        \caption{Listed-price exposure across 69 attack and nine benign API sessions. Gray crosses mark benign sessions and brick-red circles mark attack sessions; deterministic horizontal displacement resolves overlap without changing vertical positions. Hollow diamonds and bars show medians and IQRs; the dashed line and shading mark the \$0.10 moderate D4 boundary and exposures above it.}
        \label{fig:priced_exposure}
        \end{figure}

        \textbf{Priced attack sessions reach \$3.404, with 51/69 crossing the \$0.10 circuit-breaker setting.} The largest benign session in the priced cohort reaches \$0.025, and every attack vector contains at least one session above that maximum. We compute these values by applying the canonical provider-listed rate table to provider-reported input- and output-token counts (Table~\ref{tab:budget_risk}; Figure~\ref{fig:priced_exposure}).

        \begin{table}[t]
\centering
\footnotesize
\caption{Listed-price exposure in the 256-execution evaluation cohort.}
\label{tab:budget_risk}
\renewcommand{\arraystretch}{1.06}
\setlength{\tabcolsep}{2.2pt}
\begin{tabular*}{\columnwidth}{@{\extracolsep{\fill}}lrrrrr@{}}
\toprule
\raisebox{0.5\normalbaselineskip}{Cohort} & \shortstack{Priced /\\total} & \shortstack{Mean\\(USD)} & \shortstack{Median\\(USD)} & \shortstack{Max\\(USD)} & \shortstack{$>\$0.10$\\($n/N$)} \\
\midrule
Benign & 9 / 35 & \$0.014 & \$0.014 & \$0.025 & 0/9 \\
\midrule
V0 Direct & 16 / 54 & \$1.434 & \$1.293 & \$3.404 & 14/16 \\
V1 Poly. & 14 / 42 & \$1.182 & \$1.074 & \$2.994 & 12/14 \\
V2 Cross & 3 / 18 & \$0.475 & \$0.359 & \$1.066 & 2/3 \\
V3 Stealth & 19 / 48 & \$0.526 & \$0.582 & \$0.715 & 17/19 \\
V4 Weapon. & 3 / 19 & \$0.290 & \$0.062 & \$0.808 & 1/3 \\
V5 Scheduled & 14 / 40 & \$0.342 & \$0.083 & \$1.368 & 5/14 \\
\midrule
\textbf{All attacks} & 69 / 221 & \$0.820 & \$0.715 & \$3.404 & \textbf{51/69} \\
\bottomrule
\end{tabular*}
\par\vspace{\abovecaptionskip}
\begin{minipage}{\linewidth}
\raggedright
\emph{Note.} Priced/total reports executions with a positive listed-price cost over all cohort executions. Mean, median, maximum, and $>\$0.10$ are computed over the priced subset; the final column reports $n/N$ using priced $N$. Executions without a positive listed-price cost remain in total $N$ but are excluded from dollar summaries.
\end{minipage}
\end{table}

\begin{figure}[t]
        \centering
        \includegraphics[width=\columnwidth]{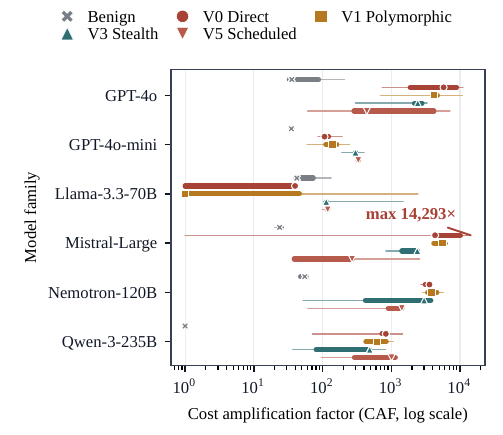}
        \caption{CAF distributions across six core model families, four primary attack vectors (V0/V1/V3/V5), and the benign comparator (206 executions; $n=5$--$11$ per condition within a family). Within each condition, thin lines span the observed minimum and maximum, thick segments denote the interquartile range, and markers denote medians. Zero-width intervals collapse to the marker. CAF is logarithmic.}
        \label{fig:caf_boxplot}
        \end{figure}

        Listed-price dollars combine repeated processing with provider-specific rates; CAF characterizes recursive input amplification independently of those rates. Figure~\ref{fig:caf_boxplot} summarizes the run-level distributions for all 206 executions and shows an orders-of-magnitude upper tail, with a maximum of 14,293$\times$. All six family-wise V5-versus-benign contrasts remain significant after Holm correction ($p=0.0155$--$0.0317$; Appendix Table~\ref{tab:caf_results}), with rank-biserial effects $r_{rb}=0.78$--$1.00$.

        Exposure also varies sharply across repeated executions. For V1 on Llama-3.3-70B, the CAF coefficient of variation is 3.08 ($n=11$).

        \noindent\begin{minipage}[t]{\columnwidth}
        \noindent\textbf{Finding 1.} Attack exposure reaches \$3.404 and 14,293$\times$ CAF, with 51/69 priced attack sessions crossing \$0.10. Observed run termination is highly variable and does not enforce the host's budget.
        \end{minipage}

        \subsection{Turn Survival Tracks Recursive Exposure}
        \label{sec:eval:alignment}

\begin{figure*}[!t]
        \centering
        \includegraphics[width=\textwidth]{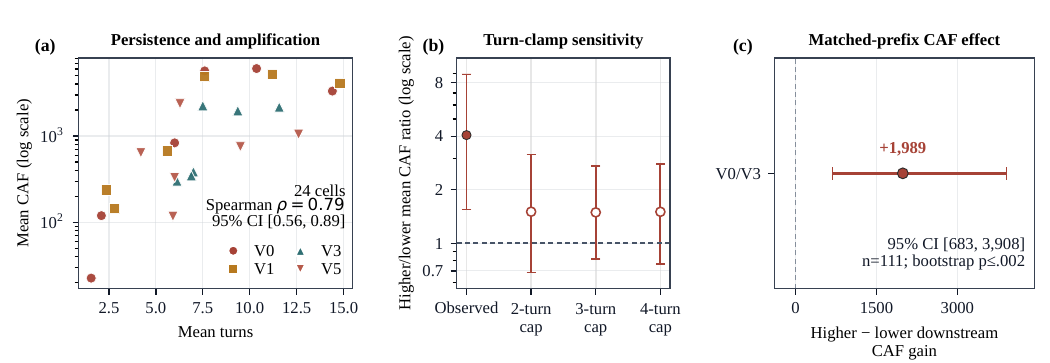}
        \caption{Turn survival tracks recursive exposure across observed deployments. (a) Mean CAF versus mean turns across the four primary vectors (V0/V1/V3/V5; 24 complete family--vector cells). (b) Ratios of mean CAF between globally ranked higher- and lower-amplification family groups, observed and after offline truncation of 44 malicious traces with recorded per-turn prompt tokens at two, three, or four turns; bars show trace-bootstrap 95\% CIs on a logarithmic axis. (c) Mean paired difference in downstream CAF gain across 111 supported V0/V3 matched prefixes (cross-fitted higher- minus lower-amplification family group); the horizontal bar gives the trace-cluster bootstrap 95\% CI (683--3,908; bootstrap $p{\leq}0.002$).}
        \label{fig:alignment_mechanism}
        \end{figure*}

        The variability in Finding 1 leaves one structural question unresolved: how much of the observed cross-deployment exposure disparity tracks turn survival after controlling for matched attack progress? We test this association at deployment and trace level (Figure~\ref{fig:alignment_mechanism}). On Llama-3.3-70B, all nine V0 Direct runs end within two turns and have a cell-mean CAF of 22.6$\times$, whereas V3 Stealth and V5 round-indexed reach 385.7$\times$ and 117.0$\times$. In this observed deployment, strategies that preserve execution realize substantially more recursive exposure than the direct payload.

        Across 24 complete cells, CAF is associated with mean turns ($\rho{=}0.79$) and tool calls ($\rho{=}0.81$). Attack-fixed and leave-one-vector-out sensitivity checks preserve the positive direction. Across 111 matched V0/V3 prefixes, the mean paired difference in downstream CAF gain between the cross-fitted higher- and lower-amplification groups is 1,989 CAF units, with a trace-cluster bootstrap 95\% CI of 683--3,908 (bootstrap $p{\leq}0.002$). Offline truncation of 44 malicious traces with recorded per-turn prompt tokens to two through four turns reduces the globally ranked higher-to-lower CAF ratio from 4.06$\times$ to 1.49--1.51$\times$. Together, these analyses quantify a strong association between turn survival and realized recursive exposure across the observed deployments.

        \subsection{Retained History Couples Continuity to Cost}

        To estimate the closed-loop effect of retaining raw history directly, we independently rerun Full, Drop, and Compress over the same 12 held-out tasks in persistent-return and matched-necessary classes on Groq and Mistral (144 planned sessions; 143 cost-complete).

\begin{figure*}[!t]
        \centering
        \includegraphics[width=\textwidth]{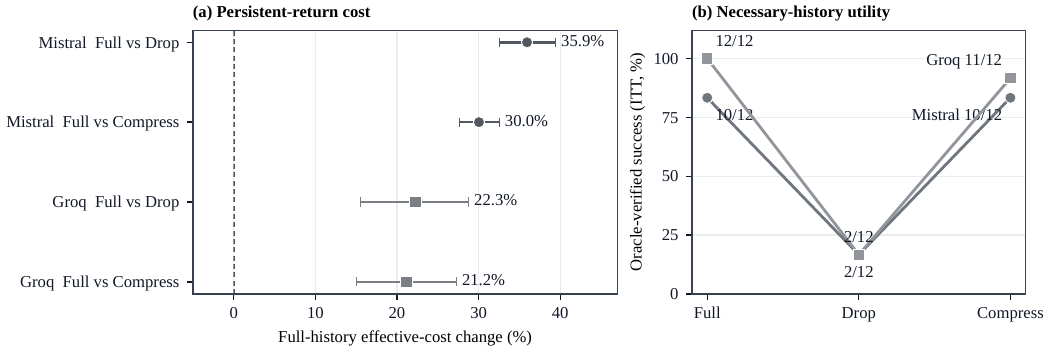}
        \caption{Raw history raises cost, whereas compression preserves most utility on tasks requiring prior history. (a) Full-over-Drop/Compress effective-cost changes on persistent returns, shown as ratios of arm means minus one with task-paired bootstrap 95\% CIs (12 planned pairs per provider--comparison); costs are complete-session totals. (b) Planned intention-to-treat (ITT) oracle-verified task success on matched-necessary tasks; non-completions remain non-successes. Both panels report the same two provider--model pairs.}
        \label{fig:history_intervention}
        \end{figure*}

        \textbf{Retaining raw history raises mean effective cost, while compression preserves necessary continuity.} On persistent-return tasks, Full costs 35.9\% more than Drop and 30.0\% more than Compress on Mistral; the corresponding increases on Groq are 22.3\% and 21.2\%. On matched-necessary tasks, Full and Compress achieve at least 10/12 successes per provider, whereas Drop records 2/12 planned/ITT successes for each provider. Appendix Table~\ref{tab:history_ablation} reports paired intervals, available-case details, and both-success counts. These independent task-paired reruns estimate the closed-loop total effect of retaining raw history relative to dropping or compressing it and show that compression preserves substantially more task utility than deletion (Figure~\ref{fig:history_intervention}).

        On a separate matched four-turn protocol, every cell completes 5/5 sessions: on Groq and Mistral, V0 raises effective cost to 1.249$\times$ and 1.308$\times$ the matched benign schedule, V5 raises it to 1.192$\times$ and 1.235$\times$, and V3 remains at 0.880$\times$ and 1.004$\times$ (Appendix Table~\ref{tab:prior_work_matched}).

        \noindent\begin{minipage}[t]{\columnwidth}
        \noindent\textbf{Finding 2.} Turn survival tracks recursive exposure; closed-loop reruns then isolate the continuity--exposure coupling: retaining raw history raises mean cost by 21.2--35.9\%, while compression preserves 10/12 and 11/12 successes on tasks requiring prior history versus 2/12 under deletion on each provider.
        \end{minipage}

        % ============================================================
        % §7 DEFENSE FRAMEWORK
        % ============================================================
        \section{Defense Framework}
        \label{sec:defense}

        The remaining defense question is whether retained state is carried into another model call. We enforce that decision at the host's pre-reingestion boundary. Retained input mass produces both absolute prompt mass and adjacent growth, mapped to D1 and D2; recursive opportunity maps to D3; generated output and price-sensitive cumulative exposure map to D4. Together, D1--D4 form a provider-agnostic host kernel that acts before the next provider call and requires no provider-private hooks. Fixed caps provide auditable containment baselines; growth-gated recursion control and host-verified progress budgets govern when legitimate workflows may receive additional opportunity or spend.

\begin{figure}[t]
        \centering
        \includegraphics[width=\columnwidth]{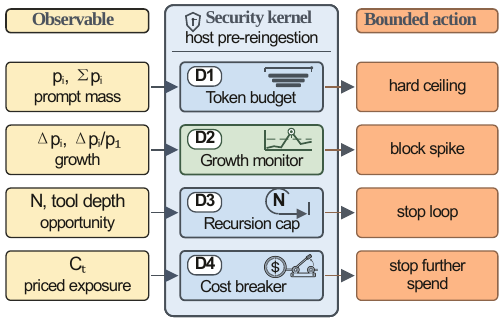}
        \caption{Four host-side invariants enforced before reingestion. D1 bounds token mass, D2 flags abnormal reingestion growth, D3 limits recursive opportunity, and D4 stops subsequent calls after the cost threshold is crossed; each gate acts before the next provider call.}
        \label{fig:defense_arch}
        \end{figure}

        Figure~\ref{fig:defense_arch} shows the resulting division of labor: D2 detects abnormal reingestion growth early. D1 and D3 bound prompt mass and recursive opportunity, while D4 stops subsequent calls once cumulative priced exposure exceeds its threshold.

        \subsection{Host-Side Enforcement Kernel}

        \noindent\textbf{D1: Token Budget Enforcement.} Per-session/per-turn envelopes bound absolute prompt-token mass. Strict, moderate, and relaxed evaluation presets use 20K/8K, 50K/20K, and 200K/100K session/per-turn limits.

        \noindent\textbf{D2: Growth-Anomaly Detection ($\Delta p$ Monitor).} Flags anomalous growth in reconstructed prompt mass. In replay and provider-telemetry checks, $\Delta p_t=p_t-p_{t-1}$ uses adjacent reported input-token counts; local integrations compute the same pre-send difference with their stated token estimator. Formally, the circuit triggers at turn $t$ if:
        \begin{equation}
        \left(\Delta p_t > \tau_{\text{abs}}\right) \vee \left(\frac{\Delta p_t}{p_1} > \theta_r \wedge \Delta p_t > \tau_{\text{min}}\right)
        \end{equation}
        where $p_1$ is first-turn prompt mass and $\tau_{\text{min}}$ suppresses tiny-seed ratios. Strict/moderate/relaxed presets use $\tau_{\text{abs}}=5/10/50$K and $\theta_r=2/3/10$, with $\tau_{\text{min}}=2$K throughout.

        \noindent\textbf{D3: Recursive-Opportunity Governance.} Frozen strict/moderate/relaxed presets cap consecutive tool turns at 3/5/10, calls per turn at 2/3/5, and total calls at 10/15/30. These caps separate parallel bursts, session-wide opportunity, and consecutive depth, providing the fixed containment baseline. The growth-gated policy conditions depth enforcement on retained-state growth; additional recursive opportunity remains subject to explicit host authorization.

        \noindent\textbf{D4: Cost Circuit Breaker.} Strict/moderate/relaxed presets terminate when cumulative priced exposure $C_t$ exceeds \$0.05/\$0.10/\$0.50, stopping subsequent calls rather than retroactively capping triggering-turn spend. This price-aware invariant covers input/output asymmetry that token-growth rules alone cannot express.

        % ============================================================
        % §8 DEFENSE EVALUATION
        % ============================================================
        \section{Defense Evaluation}
        \label{sec:defense_eval}

        A deployable kernel must stop recorded attacks, retain stopping authority under threshold-aware strategies, preserve benign and long-running work, and execute before the next provider call.

        \subsection{Frozen-Trace Containment}

        \textbf{All 123 replay evaluations of 41 recurring multi-turn attack traces are contained.} Strict, moderate, and relaxed D1--D4 terminate every trace under every preset. We include the triggering provider row in leakage; once a gate fires, no subsequent provider call is issued. Across the 80 retained trace logs (52 with turn-level records and 28 summary-only logs excluded from replay), the 41 recurring multi-turn attack traces form the replay set.

        \subsection{Threshold-Aware Layer Coverage}

        Full-stack replay establishes containment but not each gate's first-trigger coverage, because an early D2 trigger can mask the remaining gates. We therefore ablate the gates on the frozen sequences. D2 dominates: removing it delays V0/V1/V4/V5 from Turn~2 to Turns~4--6 and V3 from Turn~3 to Turn~5; leakage for those five vectors rises by 4.2--14.7$\times$, while V2 is unchanged. Removing D1, D3, or D4 leaves the first trigger unchanged on the five D2-dominated sequences.

        We address this coverage gap with constructions designed to remain below D2's growth condition and a high-growth control. The set pairs a loud D2 control with seed inflation for D1, low-and-slow, multi-tool, depth-hugging, and mixed constructions for D3, and an output-heavy construction for D4.

        \begin{figure}[t]
        \centering
        \includegraphics[width=\columnwidth]{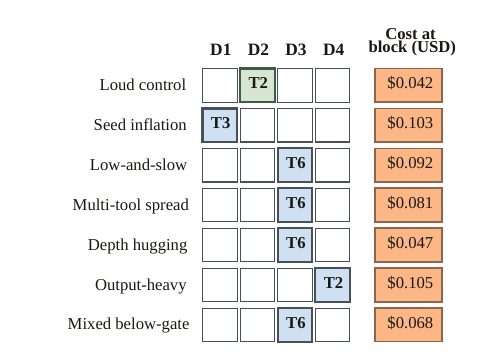}
        \caption{Seven constructed, white-box threshold-aware traces evaluated offline against the frozen moderate kernel collectively exercise all four gate conditions. Cell position identifies the first reported layer under the kernel's evaluation order, labels give the block turn, and the rightmost column reports cumulative nominal cost through and including that turn.}
        \label{fig:layer_coverage}
        \end{figure}

        The seed-inflation construction holds $\Delta p=9{,}999$ and $\Delta p/p_1<3$, remaining below D2 while crossing D1's per-turn limit at Turn~3. That turn also crosses D4's \$0.10 threshold; the ordered kernel reports D1 first at \$0.103 of trigger-inclusive leakage, while disabling D1 leaves D4 to block the same turn at the same leakage. Across the seven full-stack traces in Figure~\ref{fig:layer_coverage}, maximum observed trigger-inclusive leakage is \$0.105, caused by output-heavy turn-boundary overshoot around D4's threshold. Disabling D2 on the loud-control trace raises leakage to \$0.127 before D4 fires. Together, these constructions exercise every gate condition and show how low-growth strategies move enforcement beyond D2.

        \subsection{D2 Operating Point}
        \label{sec:eval:fpr}

        Because D2 is the dominant early detector, a viable operating point must distinguish attack growth from benign prompt evolution while keeping the full kernel inexpensive to execute. \textbf{Moderate D2 triggers on 2/1,050 benign provider--scenario executions} across five complete corpus runs (0.19\%); the primary corpus has 0/210 triggers (rule-of-three upper bound 1.4\%). Both are provider-specific small-seed outliers. A post hoc sensitivity analysis with a denominator floor removes both while shifting mean attack leakage from 4.13\% to 5.05\%. We retain the raw moderate rule as the frozen detector.

        Compared with the denominator-floor D2 sensitivity rule, a 20K static ceiling also blocks the six-vector replay set with 0/1,050 benign triggers, but acts later (mean Turn~4.67 versus 2.33) and after 26.1\% versus 5.1\% leakage. Across 1.056M local gate evaluations, median decision-logic time is 3.3/4.4~$\mu$s for D2/the full stack; provider, network, and tokenization time lie outside this measurement.

        \subsection{Progress-Aware Workflow Utility}
        We next evaluate long-workflow utility under D3 and D4, testing whether recursive opportunity and additional spend can be conditioned, respectively, on retained-state growth and host-verified progress. Frozen moderate D3 interrupts all 60 completed 8--12-step workflows at Turn~6, motivating the growth-gated policy.

        \textbf{Replacing frozen D3 with growth-gated D3 preserves 41/41 full-stack replay containment while eliminating depth-only interruptions on the 60-workflow D3 policy-design corpus.} D1 and D2 remain the first triggers on the replay cohort (1/41 and 40/41, respectively), while growth-gated D3 causes 0/60 D3 interruptions on the D3 policy-design corpus. Growth-gated D3 removes the depth bottleneck, but unchanged D4 still stops 7/60 workflows; this remaining bottleneck motivates progress-aware spend authorization.

        Progress-authorized D4 (the checkpoint policy) retains a \$0.10 base budget and releases \$0.13/\$0.16 only after monotonic, host-verified milestones; model and tool self-reports cannot authorize additional spend. On a disjoint 12-workflow Cerebras stack, it completes 9/12 protocol tasks with 0/12 defense interruptions. This second stack exercises progress-authorized execution on Cerebras; the separately frozen Mistral suite evaluates the fixed-versus-progress-authorized policy contrast.

        \textbf{D4 policy transfer.} We freeze 24 workflows across six categories, then run fixed and progress-authorized D4 independently on one provider--model pair (Mistral/Mistral Small 4), producing 48 closed-loop sessions in alternating order. D1--D3 are disabled to isolate D4, and the two arms share no action history. A host oracle verifies ordered milestones and exact frozen answer facts; an exact match to the frozen typed answer object defines oracle-verified task success under the intention-to-treat (ITT) denominator. The experiment compares two complete authorization schedules: fixed remains at \$0.10, while progress-authorized D4 may release \$0.13 or \$0.16 after verified progress.

        \begin{figure}[t]
        \centering
        \includegraphics[width=\columnwidth]{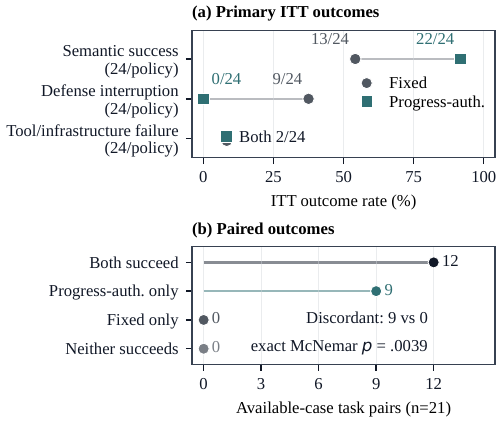}
        \caption{Mistral D4 policy-transfer outcomes. Fixed holds \$0.10; progress-authorized may release \$0.13/\$0.16. (a) ITT outcomes across 24 planned sessions per policy. (b) Available-case pairs ($n{=}21$). Exact two-sided McNemar $p{=}0.0039$ on these observed schedules.}
        \label{fig:d4_transfer}
        \end{figure}

        \begin{samepage}
        \noindent\textbf{The schedule contrast does not identify an independent benefit of progress gating.} Across 24 workflows per schedule, the checkpoint schedule completed 22/24 tasks with no pre-completion interruptions, compared with 13/24 completions and nine interruptions under fixed-\$0.10 D4. Each of those nine interruptions occurred after 9/10 verified milestones. The schedules did not have matched maximum authorized budgets: the checkpoint schedule could release \$0.13/\$0.16. Eight of the nine fixed-budget interruptions occurred when the fixed schedule ran first; in the other run position, one interruption occurred. In first-position runs, the aggregate cache-hit shares were 14.7\% for fixed D4 and 14.4\% for the checkpoint schedule. Fixed D4's second-position share was 42.7\%. These order and cache figures are descriptive. Among 21 available-case pairs, all nine discordant outcomes favor the checkpoint schedule (exact two-sided McNemar $p{=}0.0039$; Figure~\ref{fig:d4_transfer}). Mean effective cost over all sessions is \$0.113 versus \$0.099 (Appendix Table~\ref{tab:d4_transfer}). The reported contrast therefore does not separate progress gating from the higher authorized budget and the observed order and cache differences.
        \end{samepage}

        \subsection{Live Inline Placement}

        \begin{samepage}
        \noindent\textbf{D1--D4 intercept 23/25 live attack sessions before the next provider call; upstream payload limits preempt the remaining two.}
        \par
        \end{samepage}

        The suite spans five vectors and five provider--model pairs, with an upstream 413 payload cap stopping the two Groq sessions before they reach the middleware. On a separate Mistral/MCP-SDK path, the pre-reingestion loop blocks 10/10 attack-shaped sessions and completes 20/20 four-step benign tasks. Both integration paths place host enforcement before the next provider call.

        \begin{samepage}
        \noindent\textbf{Finding 3.} The host kernel stops every recurring attack trace in the frozen replay corpus, and live integrations execute the controls before the next provider call. Growth-gated D3 preserves that containment with 0/60 D3 interruptions on the D3 policy-design corpus. On the separate Mistral/MCP path, the pre-reingestion loop blocks 10/10 tested attack-shaped sessions and completes 20/20 four-step benign tasks. The unequal-budget workflow comparison does not establish an independent utility benefit from progress-gated authorization.
        \end{samepage}

        \FloatBarrier

        % ============================================================
        % §9 DISCUSSION
        % ============================================================
        \section{Discussion}
        \label{sec:discussion}

        Taken together, Findings 1--3 expose a lifecycle gap in tool security. Connecting a remote tool, or allowlisting its endpoint, authorizes that endpoint to execute. Retention can still authorize its untrusted return to occupy future context and incur future charges. Allowlisting therefore closes the question of which endpoint may be called, and leaves open the question of which returned bytes may be billed again.

        \subsection{Who Owns the Billable-State Boundary?}
        That authorization belongs to the host. Tool servers produce candidate state, agent runtimes determine how that state is retained or transformed, and providers meter the reconstructed input; only the host joins these actions with local task state, authorized budget, and the decision to issue another model call. Provider usage telemetry can support this decision: before reingestion, the host must decide whether a particular return remains necessary, how it may be represented, and whether it warrants another unit of opportunity or spend.

        \subsection{Why Semantic Safety Is Insufficient}
        Host ownership identifies the decision-maker and shifts the remaining question to the evidence required for further computation. Content safety evaluates a return's validity and acceptability; economic safety determines whether that return merits recurring authority over future computation. Every evaluated vector remains compatible with schema validity and topical plausibility, leaving reconstruction cost to explicit economic controls. Caching and routing lower unit cost, and compression reduces retained mass; authorization determines whether an untrusted return merits another unit of work. Because task-relevant state may be necessary for completion, host-controlled history transformation preserves the facts required for continuity while removing attacker-controlled bulk before it recurs as future billable input.

        \subsection{Compositional Control across Evaluated Strategies}
        History transformation controls what task-relevant state survives, but not how many later calls or how much generated output the retained state may induce. The host therefore pairs compression with D1--D4 to bound absolute prompt mass, incremental growth, recursive opportunity, and cumulative spend. This supplies both early detection and hard containment: D2 catches rapid reconstruction growth, while D1, D3, and D4 backstop cases in which growth alone is insufficient. Growth-gated D3 conditions further recursive opportunity on retained-state behavior. Progress-authorized D4 is defined to release spend only after host-verified task progress.

        The evaluated vectors shift the first visible signal---V1 representation, V2 cross-tool recursion, V3 delayed growth, V4 output pressure, and V5 round-indexed continuation---while D1--D4 bound the host-metered resources at this boundary.

        \subsection{Repository-Level Safeguard Scan}
        Of 4,535 public MCP server or transport repositories, 3,830 have eligible Python, JavaScript, TypeScript, or Go code in the collected files (median one file). Two authors independently labeled every lexical match against a written codebook, resolving disagreements by discussion (Table~\ref{tab:scan_review}). The strict scan flags 71 repositories at 75\% precision (51 of 68 decidable; Cohen's $\kappa=0.84$), and a relaxed sweep adds 20 ($\kappa=0.92$): 71 (1.9\%) enforce a verified safeguard in collected files. Because these files omit most code, we audited 100 random repositories in full at the scanned commits (95 still public; median 26 code files), labeling new matches alike ($\kappa=0.86$). The full-code audit found verified safeguards in 17 of 95 repositories (17.9\%; Wilson 95\% CI 11.5--26.8\%). One repository bounds cost, and none combines all four families. Sixteen of these 17 repositories were missed when only the collected files were scanned, showing that the collected-file result understates coverage within this audited sample. Because the audit inspects source code at the scanned commits, it does not estimate controls implemented only in deployment configuration or controls that are not expressed lexically.

        % ============================================================
        % §10 CONCLUSION
        % ============================================================
        \begin{samepage}
        \section{Conclusion}
        \label{sec:conclusion}

        Denial-of-Wallet is a lifecycle failure: admission permits a tool to run, but retention can grant its untrusted return recurring economic authority. Our measurements expose recurring cost. The independent retention intervention isolates the contribution of raw history, while compression preserves history-dependent continuation. Controls at the pre-reingestion boundary bound prompt mass, context growth, recursive opportunity, and cumulative spend. The ecosystem scan shows that verified code-visible safeguards remain uncommon. Persistent billable state is therefore a first-class security object: the decisive question is what a tool's return is authorized to become---future spend.
        \end{samepage}

        \appendices

        % ============================================================
        % APPENDIX A: ETHICAL CONSIDERATIONS
        % ============================================================
        \section{Ethical Considerations}
        \label{app:ethics}

        Live experiments used author-controlled API accounts and tool endpoints under explicit turn and budget caps. No human subjects were involved, and no personal data were collected. We privately disclosed the lifecycle risk, reproduction conditions, and D1--D4 mitigations through official security channels to Mistral, Groq, LangChain, LlamaIndex, and the MCP Python SDK maintainers. The MCP Python SDK maintainers closed their report, stating that limiting how tool results are fed back into model context is the host application's job and that the SDK does not re-ingest tool results. The other reports remain under vendor or maintainer review. Operational attack material remains withheld from public release during coordinated disclosure. We publish the problem definition, measurements, host-side controls, and reproducible replay evidence because independent rediscovery is plausible and the defensive value outweighs residual misuse risk under coordinated release. We do not authorize any application of this research to illegal attacks or any other illegal use in the real world.

        % ============================================================
        % REFERENCES
        % ============================================================
        \bibliographystyle{IEEEtran}
        \bibliography{references}
        % ============================================================
        % APPENDIX C: SUPPLEMENTARY TABLES
        % Continues in the column after the last reference.
        % ============================================================
        \section{Supplementary Tables}
        \label{app:tables}

        \makeatletter
        \begin{center}
        \footnotesize
        \def\@captype{table}
        \caption{D4-only Mistral Small 4 transfer: 24 frozen workflows independently executed under each policy (48 sessions).}
        \label{tab:d4_transfer}
        \renewcommand{\arraystretch}{1.08}
        \begin{tabular}{@{}lcc@{}}
        \toprule
        \textbf{Metric} & \textbf{Fixed} & \textbf{Progress-auth.} \\
        \midrule
        ITT task success & 13/24 (54.2\%) & 22/24 (91.7\%) \\
        Pre-completion D4 stop & 9/24 & 0/24 \\
        Tool failure & 2/24 & 2/24 \\
        \midrule
        Mean cost/session (USD) & 0.0992 & 0.1126 \\
        Median cost/session (USD) & 0.1037 & 0.1143 \\
        \bottomrule
        \end{tabular}
        \par\vspace{\abovecaptionskip}
        \begin{minipage}{\columnwidth}
        \footnotesize\raggedright
        \emph{Note.} Fixed holds D4 at \$0.10; progress-authorized D4 releases \$0.13/\$0.16 at 7/10 and 10/10 host-verified milestones. ITT success requires all 10 milestones and an exact match to the frozen typed answer object. Cost rows report nominal cache-aware effective cost in USD per session over all 24 sessions per policy.
        \end{minipage}
        \end{center}
        \makeatother

        \makeatletter
        \setlength{\@dblfptop}{0pt}
        \setlength{\@dblfpbot}{0pt plus 1fil}
        \makeatother
        \begin{table*}[t]
        \centering
        \footnotesize
        \setlength{\tabcolsep}{3pt}
        \caption{Closed-loop history-policy cost effects.}
        \label{tab:history_ablation}
        \begin{tabular*}{\textwidth}{@{\extracolsep{\fill}}llrrrllr@{}}
\toprule
\textbf{Return} & \textbf{Comparison} & \textbf{$n$} & \textbf{$m$} & \textbf{$\Delta$ input} & \textbf{$\Delta$ cost [95\% CI]} & \textbf{Median $\Delta\$$ [IQR]} & \textbf{Both-success ratio} \\
\midrule
\multicolumn{8}{@{}l}{\textit{Groq}} \\
\textbf{Persistent} & Full/Drop & 12/12 & 10 & 37.88\% & \textbf{22.26\%} [15.56, 28.76] & \$0.00012 [\$0.00008, \$0.00014] & 1.25$\times$ \\
\textbf{Persistent} & Full/Compress & 12/12 & 10 & 33.10\% & \textbf{21.15\%} [15.04, 27.27] & \$0.00011 [\$0.00007, \$0.00013] & 1.22$\times$ \\
Necessary & Full/Drop & 11/12 & 2 & 35.50\% & 22.13\% [0.46, 42.20] & \$0.00014 [\$0.00009, \$0.00020] & 1.54$\times$ \\
Necessary & Full/Compress & 12/12 & 11 & 31.38\% & 29.15\% [20.26, 38.57] & \$0.00016 [\$0.00013, \$0.00019] & 1.30$\times$ \\
\addlinespace[1pt]
\multicolumn{8}{@{}l}{\textit{Mistral}} \\
\textbf{Persistent} & Full/Drop & 12/12 & 10 & 45.97\% & \textbf{35.93\%} [32.50, 39.42] & \$0.00010 [\$0.00010, \$0.00012] & 1.37$\times$ \\
\textbf{Persistent} & Full/Compress & 12/12 & 10 & 40.30\% & \textbf{30.05\%} [27.65, 32.53] & \$0.00008 [\$0.00008, \$0.00009] & 1.31$\times$ \\
Necessary & Full/Drop & 12/12 & 2 & 51.60\% & 38.88\% [36.51, 41.24] & \$0.00012 [\$0.00012, \$0.00012] & 1.39$\times$ \\
Necessary & Full/Compress & 12/12 & 10 & 46.18\% & 33.03\% [29.83, 36.09] & \$0.00011 [\$0.00010, \$0.00012] & 1.32$\times$ \\
\bottomrule
\end{tabular*}

        \par\vspace{\abovecaptionskip}
        \begin{minipage}{\textwidth}
        \footnotesize\raggedright
        \emph{Note.} Persistent-return rows are the primary cost estimand; matched-necessary rows assess continuity when returns carry task-required records. $n$ gives evaluable/planned pairs, and $m$ gives both-success pairs. $\Delta$ input and $\Delta$ cost are Full/comparator ratios of arm means minus one; 95\% CIs use 20,000 task-paired percentile-bootstrap resamples. Median $\Delta\$$ is Full minus comparator. The Groq Necessary--Drop comparison uses 11 available pairs after one parse failure. Both-success ratios with $m=2$ are descriptive summaries.
        \end{minipage}
        \end{table*}

        \begin{table*}[t]
        \centering
        \footnotesize
        \setlength{\tabcolsep}{3pt}
        \caption{Author review of the census and full-code audit; counts are verified\,/\,not a safeguard\,/\,undecided.}
        \label{tab:scan_review}
        \renewcommand{\arraystretch}{1.04}
        \begin{tabularx}{\textwidth}{@{}>{\raggedright\arraybackslash}p{0.084\textwidth}>{\raggedright\arraybackslash}X>{\centering\arraybackslash}p{0.072\textwidth}>{\raggedright\arraybackslash}p{0.155\textwidth}>{\centering\arraybackslash}p{0.078\textwidth}>{\centering\arraybackslash}p{0.064\textwidth}>{\centering\arraybackslash}p{0.074\textwidth}>{\centering\arraybackslash}p{0.068\textwidth}@{}}
        \toprule
        \textbf{Family} & \textbf{Census strict scan: example patterns} & \textbf{Matches} & \textbf{Census relaxed sweep: example patterns} & \textbf{Candidates} & \textbf{Verified repos} & \textbf{Full-code matches} & \textbf{Full-code verified} \\
        \midrule
        D1 mass & \texttt{max\_\{input,\allowbreak prompt,\allowbreak context\}\_tokens}; \texttt{token\_budget}; \texttt{token\_limit} near context or tool & 12 (7/4/1) & \texttt{max\_message\_\allowbreak tokens}; input or context token cap & 0 & 7 & 7 (5/2/0) & 5 \\
        D2 rate & \texttt{rate\_limiter}; \texttt{window\_ms}; token or leaky bucket; sliding or rolling window near rate limit & 23 (17/6/0) & \texttt{rate\_limit}; throttle; request quota & 93 (18/73/2) & 35 & 35 (6/28/1) & 7 \\
        D3 recursion & \texttt{max\_\{iterations,\allowbreak steps,turns,\allowbreak tool\_calls\}}; \texttt{loop\_limit}; \texttt{max\_consecutive}; visited states & 39 (29/8/2) & \texttt{recursion\_limit}; \texttt{max\_\{turn,step\}} & 4 (2/2/0) & 31 & 22 (9/12/1) & 9 \\
        D4 spend & \texttt{max\_cost}; \texttt{budget\_usd}; cost limit or budget; spend limit; cost, spend, or USD near cap or guard & 2 (2/0/0) & cost, spend, USD, or billing near budget, limit, or cap & 14 (0/14/0) & 2 & 9 (1/8/0) & 1 \\
        \midrule
        Repositories & 71 of 3,830 flagged & 51/17/3 & 105 of the other 3,759 & 20/83/2 & 71 & 17/27/0 & 17 of 95 \\
        \bottomrule
        \end{tabularx}
        \par\vspace{\abovecaptionskip}
        \begin{minipage}{\textwidth}
        \footnotesize\raggedright
        \emph{Note.} All passes skip test, documentation, and vendored paths, apply the same noise filters, and keep one match per family per repository. The audit scans all code; census-reviewed matches keep their labels. Two authors labeled all items independently, resolving 13 disagreements by discussion; undecided repositories are no longer public.
        \end{minipage}
        \end{table*}
        \begin{table*}[t]
        \centering
        \footnotesize
        \setlength{\tabcolsep}{4pt}
        \caption{Success-gated retention-cost comparison of mechanism-matched local schedules on matched four-turn workflows.}
        \label{tab:prior_work_matched}
        \renewcommand{\arraystretch}{1.05}
\begin{tabular}{@{}lccc@{}}
\toprule
\multirow[c]{2}{*}[-6pt]{Treatment} & \multirow[c]{2}{*}[-6pt]{\shortstack{Task\\success}} & \multicolumn{2}{c}{Cost ratio} \\
\cmidrule(lr){3-4}
& & \multicolumn{1}{c}{\shortstack{Effective /\\benign}} & \multicolumn{1}{c}{\shortstack{Gross /\\projected}} \\
\midrule
\multicolumn{4}{@{}l}{\textit{Groq}} \\
V0 & 5/5 & 1.249$\times$ & 1.675$\times$ \\
V3 & 5/5 & 0.880$\times$ & 1.344$\times$ \\
V5 & 5/5 & 1.192$\times$ & 1.593$\times$ \\
Matched benign & 5/5 & 1.000$\times$ & 1.555$\times$ \\
\addlinespace[3pt]
\multicolumn{4}{@{}l}{\textit{Mistral}} \\
V0 & 5/5 & 1.308$\times$ & 1.786$\times$ \\
V3 & 5/5 & 1.004$\times$ & 1.386$\times$ \\
V5 & 5/5 & 1.235$\times$ & 1.690$\times$ \\
Matched benign & 5/5 & 1.000$\times$ & 1.660$\times$ \\
\bottomrule
\end{tabular}

        \par\vspace{\abovecaptionskip}
        \begin{minipage}{\textwidth}
        \footnotesize\raggedright
        \emph{Note.} Each provider--treatment cell contains five sessions; task success requires the four-turn/four-tool-call oracle. Ratios use mean per-session costs over successful sessions. Effective/benign compares observed cache-aware effective cost with the matched benign schedule. Gross/projected divides observed gross cost by a fixed-action no-history projection over the same tool-call and completion-token sequence. V0, V3, and V5 are local schedules on this protocol. The table does not rank them against Overthinking Loops or Beyond Max Tokens.
        \end{minipage}
        \end{table*}

        % Auto-generated from the canonical database view. Do not edit by hand.
\begin{table*}[!t]
\centering
\caption{CAF distributions across six core LLM families, four primary attack vectors, and the benign comparator.}
\label{tab:caf_results}
\resizebox{\textwidth}{!}{
\begin{tabular}{llrrrrrc}
\toprule
\raisebox{0.5\normalbaselineskip}{Model} & \raisebox{0.5\normalbaselineskip}{Metric} & \raisebox{0.5\normalbaselineskip}{Benign} & \raisebox{0.5\normalbaselineskip}{Direct (V0)} & \raisebox{0.5\normalbaselineskip}{Polymorphic (V1)} & \raisebox{0.5\normalbaselineskip}{Stealth (V3)} & \raisebox{0.5\normalbaselineskip}{Scheduled (V5)} & \shortstack{V5 vs.\ benign\\adj.\ $p$ / $r_{rb}$} \\
\midrule
\multirow{4}{*}{\textbf{GPT-4o}} & Median CAF & 35.6 & 5,778.3 & 4,204.9 & 2,427.8 & 440.2 & \multirow{4}{*}{\shortstack{$0.0317^{*}$\\$r_{rb}=0.89$}} \\
 & IQR & 32.8--87.5 & 1,889.8--8,921.7 & 3,997.1--4,692.2 & 2,177.7--2,798.8 & 286.5--4,169.9 &  \\
 & Max & 208.7 & 11,116.0 & 10,820.5 & 3,328.6 & 7,143.9 &  \\
 & Valid $n$ & 5 & 5 & 5 & 6 & 7 &  \\
\midrule
\multirow{4}{*}{\textbf{GPT-4o-mini}} & Median CAF & 35.2 & 106.7 & 139.9 & 301.6 & 331.1 & \multirow{4}{*}{\shortstack{$0.0317^{*}$\\$r_{rb}=1.00$}} \\
 & IQR & 35.1--35.5 & 106.7--122.3 & 111.3--161.3 & 301.5--303.4 & 330.9--331.2 &  \\
 & Max & 37.8 & 192.8 & 247.0 & 402.1 & 331.3 &  \\
 & Valid $n$ & 5 & 10 & 5 & 8 & 5 &  \\
\midrule
\multirow{4}{*}{\textbf{Llama-3.3-70B}} & Median CAF & 42.3 & 39.8 & 1.0 & 113.2 & 118.8 & \multirow{4}{*}{\shortstack{$0.0155^{*}$\\$r_{rb}=0.78$}} \\
 & IQR & 41.4--73.3 & 1.0--39.8 & 1.0--46.0 & 113.2--113.2 & 118.7--118.8 &  \\
 & Max & 133.2 & 39.8 & 2,436.5 & 1,486.6 & 118.9 &  \\
 & Valid $n$ & 10 & 9 & 11 & 5 & 11 &  \\
\midrule
\multirow{4}{*}{\textbf{Mistral-Large}} & Median CAF & 23.7 & 4,336.6 & 5,561.1 & 2,394.3 & 269.2 & \multirow{4}{*}{\shortstack{$0.0317^{*}$\\$r_{rb}=1.00$}} \\
 & IQR & 23.2--25.1 & 4,336.6--10,136.3 & 4,167.1--6,108.3 & 1,423.7--2,394.3 & 38.7--269.3 &  \\
 & Max & 26.4 & 14,293.4 & 6,277.1 & 2,394.3 & 2,588.3 &  \\
 & Valid $n$ & 5 & 11 & 5 & 11 & 5 &  \\
\midrule
\multirow{4}{*}{\textbf{Nemotron-120B}} & Median CAF & 55.1 & 3,599.8 & 3,847.3 & 3,007.2 & 1,426.2 & \multirow{4}{*}{\shortstack{$0.0317^{*}$\\$r_{rb}=1.00$}} \\
 & IQR & 47.9--57.7 & 3,120.3--3,599.9 & 3,380.7--4,612.2 & 420.9--3,761.4 & 905.0--1,426.5 &  \\
 & Max & 59.3 & 3,602.0 & 5,750.2 & 3,772.7 & 1,426.6 &  \\
 & Valid $n$ & 5 & 5 & 5 & 7 & 5 &  \\
\midrule
\multirow{4}{*}{\textbf{Qwen-3-235B}} & Median CAF & 1.0 & 833.9 & 616.9 & 485.2 & 1,010.7 & \multirow{4}{*}{\shortstack{$0.0269^{*}$\\$r_{rb}=1.00$}} \\
 & IQR & 1.0--1.0 & 734.1--833.9 & 428.0--843.4 & 80.7--485.2 & 291.7--1,144.7 &  \\
 & Max & 1.0 & 1,460.5 & 1,064.5 & 830.8 & 1,144.8 &  \\
 & Valid $n$ & 5 & 10 & 5 & 9 & 6 &  \\
\bottomrule
\end{tabular}}
\par\vspace{\abovecaptionskip}
\begin{minipage}{\textwidth}
\footnotesize\raggedright
\emph{Note.} The final column reports two-sided Mann--Whitney $p$-values Holm-adjusted across the six family-wise comparisons and rank-biserial effects $r_{rb}$ for V5 versus benign. Positive $r_{rb}$ indicates higher CAF under V5 (* $p<0.05$, ** $p<0.01$).
\end{minipage}
\end{table*}

        \begin{table*}[!t]
        \centering
        \footnotesize
        \caption{Construction and run settings of the benign, workflow, and history-policy corpora.}
        \label{tab:protocols}
        \renewcommand{\arraystretch}{1.0}
        \setlength{\tabcolsep}{3pt}
        \begin{tabularx}{\textwidth}{@{}>{\raggedright\arraybackslash}p{0.13\textwidth}>{\raggedright\arraybackslash}X>{\raggedright\arraybackslash}p{0.19\textwidth}>{\raggedright\arraybackslash}p{0.19\textwidth}>{\raggedright\arraybackslash}p{0.18\textwidth}@{}}
        \toprule
        \textbf{Corpus} & \textbf{Construction} & \textbf{Separation and freezing} & \textbf{Run settings} & \textbf{Recovery and reruns} \\
        \midrule
        \textbf{Benign operating point}: 210 scenarios on five provider--model pairs & Author-written prompts, 30 in each of seven categories: single-tool query, multi-step retrieval, code generation and debugging, data analysis, multi-tool orchestration, document exploration, and complex reasoning; simulated tool returns. & Same prompts on every pair; D2 applied offline to recorded per-turn input. & At most eight turns; 1,024 output tokens; automatic tool choice; provider-default temperature. & Up to five attempts on rate-limit, server, timeout, or connection errors. \\
        \textbf{D3 policy design}: 20 designs $\times$ three runs & Author-written packet workflows, four in each of five categories: data reconciliation, iterative debugging, policy review, project planning, and research synthesis; one 8-step and two 12-step runs per design. & Used to design growth-gated D3; the 12 held-out workflows below are disjoint. & Cerebras GPT-OSS-120B; automatic tool choice; 768 output tokens; at most 10 or 16 turns; provider-default temperature. & 12-step runs re-prompt after a missing tool call within the turn limit. \\
        \textbf{Progress-authorized held-out}: 12 workflows & Two in each of six categories: incident postmortem, control mapping, scientific protocol review, dependency migration, capacity planning, and vendor assessment; 12 host-verified milestones. & Disjoint from the 20 designs; frozen before the live run, with no held-out result used before the freeze. & Cerebras GPT-OSS-120B; temperature 0; 256 output tokens per call; at most 15 turns. & One corrective prompt after an invalid or missing call; only infrastructure-failed scenarios rerun. \\
        \textbf{History intervention}: 12 held-out tasks & Two in each of six operations: sum, weighted checksum, ordered concatenation, maximum-score label, ascending sort, and majority vote; three tool steps with 2,048-byte packets; task success by the exact host oracle. & Disjoint from two calibration tasks; held-out runs opened on protocol checks, not on observed cost direction. & Groq GPT-OSS-120B and Mistral Small 4; temperature 0; 512 output tokens. & Only sessions recorded as infrastructure failures rerun, with tasks, policies, and model unchanged. \\
        \textbf{D4 policy transfer}: 24 workflows $\times$ two policies & Four in each of six categories; 10 sequential milestones over 18,000-token packets drawn from 90 frozen repository files; typed answer object. & Frozen before the first formal provider call; each policy runs first in 12 of 24 pairs. & Mistral Small 4; temperature 0; required tool choice; no parallel calls; 256 final-output tokens; no prompt-cache key. & One corrective prompt per session; only HTTP 429 responses retried; no session reruns; every outcome reported. \\
        \bottomrule
        \end{tabularx}
        \par\vspace{\abovecaptionskip}
        \begin{minipage}{\textwidth}
        \footnotesize\raggedright
        \emph{Note.} The historical 243-execution core is not listed; its manifests do not record decoding or turn limits.
        \end{minipage}
        \end{table*}

        \FloatBarrier

        \end{document}